\documentclass[ste,twoside]{stefano}

\usepackage{amsmath}
\usepackage{amssymb}
\usepackage{graphicx}
\usepackage{epsfig}
\usepackage{rotating}
\usepackage{cite}
\usepackage{color}

\def\makeheadbox{{%
\hbox to0pt{\vbox{\baselineskip=10dd\hrule\hbox
to\hsize{\vrule\kern3pt\vbox{\kern3pt \hbox{  {\sc Eur. Phys. J. C
{\bf 37}, 471-480  (2004)} } \hbox{ {\sc
{\color{blue}{dma}}[{\color{black}{imecc}}]{\color{red}{UniCamp}}
} \hspace*{10.3cm} {\color{blue}{$\boldsymbol{\Sigma \delta
\Lambda}$}} }
\kern3pt}\hfil\kern3pt\vrule}\hrule}%
\hss}}}

\def\0{\mbox{\tiny $0$}}
\def\1{\mbox{\tiny $1$}}
\def\2{\mbox{\tiny $2$}}
\def\3{\mbox{\tiny $3$}}
\def\4{\mbox{\tiny $4$}}
\def\5{\mbox{\tiny $5$}}
\def\6{\mbox{\tiny $6$}}
\def\7{\mbox{\tiny $7$}}
\def\8{\mbox{\tiny $8$}}
\def\9{\mbox{\tiny $9$}}

\def\scalar{\mbox{\tiny $scalar$}}
\def\Dirac{\mbox{\tiny $Dirac$}}

\def\f14{\mbox{\tiny $\frac{1}{4}$}}
\def\infm{\mbox{\tiny $-\infty$}}
\def\infp{\mbox{\tiny $+\infty$}}

\def\min{\mbox{\small $-$}}
\def\mi{\mbox{\tiny $-$}}
\def\pl{\mbox{\tiny $+$}}
\def\ppm{\mbox{\tiny $\pm$}}

\begin{document}
%

\title{DIRAC SPINORS AND FLAVOR OSCILLATIONS}

\author{
Alex E. Bernardini\inst{1}
\and Stefano De Leo\inst{2}
}

\institute{
Department of Cosmic Rays and Chronology, State University of Campinas,\\
PO Box 6165, SP 13083-970, Campinas, Brazil,\\
{\em alexeb@ifi.unicamp.br} \and
Department of Applied Mathematics, State University of Campinas,\\
PO Box 6065, SP 13083-970, Campinas, Brazil,\\
{\em deleo@ime.unicamp.br}
}


\date{{\em June, 2004}}

\abstract{In the standard treatment of particle oscillations the
mass eigenstates are implicitly assumed to be {\em scalars} and,
consequently, the spinorial form of neutrino wave functions is
{\em not} included in the calculations. To analyze this additional
effect, we discuss the oscillation probability formula obtained by
using the Dirac equation as evolution equation for the neutrino
mass eigenstates. The initial localization of the spinor state
also implies an interference between positive and negative energy
components of mass eigenstate wave packets which modifies the
standard oscillation probability.}









\titlerunning{Dirac spinors and flavor oscillations}

\maketitle


\section{Introduction}
\label{intro}
Since a long time, particle mixing \cite{Gel55} and
oscillation \cite{Pic55,Gri69} continue
to stimulate interesting and sometimes fascinating discussions on
the many subtleties of quantum mechanics involved in oscillation
phenomena.
Measurements of various features of the fluxes of atmospheric \cite{Egu03}
and solar \cite{Fuk02,Ahm02} neutrinos
have provided, in the last years, evidence for neutrino
oscillations and therefore for neutrino masses
and mixing.
In particular, it renewed the interest in understanding the derivation
of the flavor conversion probability formula and in overcoming the
main difficulties hidden in the standard theoretical approaches.
In particular, an increasing number of theoretical papers have
recently questioned the validity of the standard plane wave
treatment of oscillations by resorting to {\em intermediate}
\cite{Giu98,Zra98,DeL04} and {\em external} \cite{Giu02B,Beu03} wave
packet frameworks.

The standard plane wave treatment \cite{Kay89,Kay02} is certainly
the simplest and probably the most intuitive way to introduce the
oscillation length and to immediately obtain an expression for
the oscillation probability. In such a formalism, a plane wave is
associated with each mass eigenstate. For the two-flavor case, the
mass eigenstate phase difference is
\begin{equation}
\label{00} \Delta \Phi = \Delta \, (E\,T-p\,L)~.
\end{equation}
Thus, an initially \textit{pure} flavor-eigenstate will be modified with
time and distance. The probability for flavor transition is
usually expressed in terms of the mixing angle $\theta$ and of the
relative phase $\Delta \Phi$ by
\begin{equation}
P(\mbox{\boldmath$\nu_\alpha$}\rightarrow\mbox{\boldmath$\nu_\beta$})
 = \sin^{\2} [2 \theta] \, \sin^{\2} \left[\frac{\Delta \Phi}{2}\right]~.
\end{equation}
The Lorentz invariant difference of phase $\Delta \Phi$  is then
conventionally evaluated by setting $\Delta T=\Delta L = 0$ and
considering, for ultra-relativistic particles, $T\approx L$ and
$p_{\1,\2} \approx E_{\1,\2}$, i.e.
\begin{equation}
\label{000}
 \Delta \Phi =   T \, \Delta E - L \, \Delta p \approx L
\, \left( \Delta E - \Delta p \right)
\approx  \frac{\Delta m^{\2}}{2 \bar{p}} \, L.
\end{equation}
By using such an approximation, one gets the well-known expression \cite{Kay02}
\begin{equation}
\label{0000}
P(\mbox{\boldmath$\nu_\alpha$}
\rightarrow\mbox{\boldmath$\nu_\beta$};L) = \sin^{\2} [2 \theta]
\, \sin^{\2} \left[ \frac{\Delta m^{\2}}{4 \bar{p}} \, L \right]~.
\end{equation}
In the plane wave formalism, the most controversial point is
certainly represented by the derivation of formulas containing
extra factors in the oscillation length
\cite{Fie03,Giu02A,Giu01,Tak01,DeL00}.
The use of wave packets allows us to
understand the origin of these extra factors.
In the plane wave approach, it is implicitly assumed that at
creation the flavor-eigenstate is unique even up to the phase at
all points and times of creation. In the wave packet treatment, at
time $T$ and at a fixed position in the overlapping region, one
experiences the interference between space points whose separation
at creation is given by $\Delta v \, T$ and this implies that an
additional initial phase is automatically included in the wave
packet formalism \cite{Giu98,DeL04}. The final result contains the
difference of phase given in Eq.(\ref{000}).
We do not intend here to re-discuss the many
controversies in the plane wave derivations of the oscillation
probability formula. We only remark that a plane wave approach
leads to conceptual difficulties and fails to explain fundamental
aspects of particle oscillations (i.e.
localization and coherence length). Wave packets eliminates some
of these problems \cite{Kay81}. In fact, the use of wave packets for
propagating mass eigenstates
({\em intermediate} wave packet model) guarantees the existence of
a coherence length, avoids the ambiguous approximations in the plane
wave derivation of the phase difference and, under particular conditions of
minimal {\em slippage}
recovers the oscillation probability given in Eq.(\ref{0000}).
Unfortunately, it is not easy to determine the size of the wave
packets at creation and it is not clear whether it makes sense to consider a
unique time of creation \cite{Ric93,DeL04}. A common argument
against the {\em intermediate} wave packet formalism is that oscillating
neutrinos are neither prepared nor  observed. Consequently, it
would be more convenient to write a transition probability between
the observable particles involved in the production and detection
process. This point of view characterizes the so-called {\em external}
wave packet approach \cite{Giu02B,Beu03}. The oscillating
particle, described as an internal line of a Feynman diagram by a
relativistic mixed scalar propagator, propagates between the
source and target ({\em external}) particles represented by wave
packets.
The function which represents the overlap of
the incoming and outgoing wave packets
in the {\em external} wave packet model corresponds to
the wave function of the
propagating mass eigenstate in the {\em intermediate} wave packet formalism.
Remarkably, it could be shown that the probability densities for
ultra-relativistic stable oscillating particles
in both frameworks are mathematically equivalent \cite{Beu03}.
However, the {\em intermediate} wave packet picture brings up a
problem, as the overlap function takes into account not only the
properties of the source, but also of the detector. This is
unusual for a wave packet interpretation and not satisfying for
causality \cite{Beu03}. This point was clarified by Giunti
\cite{Giu02B} who solves this problem by proposing an improved
version of the {\em intermediate} wave packet model where the wave
packet of the oscillating particle is explicitly computed with
field-theoretical methods in terms of {\em external} wave packets.
Despite not being applied in a completely free way, the ({\em
intermediate}) wave packet treatment commonly simplifies the
discussion of some physical aspects going with the oscillation
phenomena \cite{DeL04,Tak01}. Thus, it makes sense, as a
preliminary investigation, to consider a wave packet associated
with the propagating particle.

In this paper, we aim to investigate how the oscillation formula
is modified by using {\em fermionic} instead of {\em scalar}
particles. To do it, we shall use the Dirac equation as the
evolution equation for the mass eigenstates. Before introducing
the Dirac formalism, in section II, we briefly review the {\em
intermediate} wave packet model for scalar particles \cite{Kay81}.
In this section, by choosing a {\em gaussian} wave packet to
describe the localization of our initial flavor state, we  obtain
an analytical expression for the flavor conversion probability.
This allows us to identify the wave packet {\em slippage} and {\em
spreading} effects. In section III, we introduce the Dirac
formalism and show that a superposition of both positive and
negative frequency solutions of Dirac equation is often a
necessary condition to correctly describe the time evolution of
the mass eigenstate wave packets. We give, for strictly peaked
momentum distributions and ultra-relativistic particles, an
analytic expression for the Dirac flavor conversion probability.
The results obtained in the context of a wave packet treatment of
oscillation phenomena are (briefly) compared with quantum field
theory calculations \cite{Beu03,Giu93,Bla95}. This allow to
understand how our analysis could be included within the {\em
external} wave packets formalism. We draw our conclusions in
Section IV.

\section{Gaussian wave packets}

 The main aspects of oscillation phenomena can be understood by
studying the two flavor problem. In addition, substantial
mathematical simplification results from the assumption that the
space dependence of wave functions is one-dimensional ($z$-axis).
Therefore, we shall use these simplifications to calculate the
oscillation probabilities. In this context, the time evolution of
flavor wave packets can be described by
\begin{eqnarray}
\Phi(z,t) &=& \phi_{\1}(z,t)\cos{\theta}\,\mbox{\boldmath$\nu_{\1}$} + \phi_{\2}(z,t)\sin{\theta}\,\mbox{\boldmath$\nu_{\2}$}\nonumber\\
          &=& \left[\phi_{\1}(z,t)\cos^{\2}{\theta} + \phi_{\2}(z,t)\sin^{\2}{\theta}\right]\,\mbox{\boldmath$\nu_\alpha$} \nonumber\\
          &&~~~~~~~~~~~~+ \left[\phi_{\1}(z,t) - \phi_{\2}(z,t)\right]\cos{\theta}\sin{\theta}\,\mbox{\boldmath$\nu_\beta$}\nonumber\\
          &=& \phi_{\alpha}(z,t;\theta)\,\mbox{\boldmath$\nu_\alpha$} + \phi_{\beta}(z,t;\theta)\,\mbox{\boldmath$\nu_\beta$},
\label{0}
\end{eqnarray}
where {\boldmath$\nu_\alpha$} and {\boldmath$\nu_\beta$} are flavor-eigenstates and {\boldmath$\nu_{\1}$} and {\boldmath$\nu_{\2}$} are mass eigenstates.
The probability of finding a flavor state $\mbox{\boldmath$\nu_\beta$}$ at the instant $t$ is equal to the integrand squared modulus of the $\mbox{\boldmath$\nu_\beta$}$ coefficient
\begin{eqnarray}
P_{\scalar}(\mbox{\boldmath$\nu_\alpha$}\rightarrow\mbox{\boldmath$\nu_\beta$};t)
& = & \mbox{$ \int_{_{\infm}}^{^{\infp}} dz
\,\left|\phi_{\beta}(z,t;\theta)\right|^{\2}$} \nonumber \\
 & = &
\mbox{$\frac{\sin^{\2}{[2\theta]}}{2}\left\{\, 1 - \mbox{\sc Int}_{\scalar}(t) \, \right\}
$} ,
\label{1}
\end{eqnarray}
where $\mbox{\sc Int}_{\scalar}(t)$ represents the interference oscillating
term between the (scalar) mass eigenstate wave packets
$\phi_{\1}(z,t)$ and $\phi_{\2}(z,t)$, i.e.
\begin{equation}
\mbox{\sc Int}_{\scalar}(t) =\mbox{$ Re \left[\, \int_{_{\infm}}^{^{\infp}}dz
\,\phi^{\dagger}_{\1}(z,t) \phi_{\2}(z,t) \, \right]$}.
\end{equation}
Let us consider mass eigenstate wave packets given at time $t = 0$ by
\begin{equation}
\phi_i(z,0) = \left(\frac{2}{\pi a^{\2}}\right)^{\frac{1}{4}} \exp{\left[- \frac{z^{\2}}{a^{\2}}\right]} \exp{[i p_i \, z]}
\label{3}.
\end{equation}
The wave functions which describe their time evolution are
\begin{equation}
\phi_i(z,t) = \mbox{$\int_{_{\infm}}^{^{\infp}}\frac{dp_z}{2 \pi} \,
\varphi(p_z - p_i) \exp{\left[-i\,E_i(p_z)\,t +i \, p_z
\,z\right]}$}, \label{4}
\end{equation}
where
\begin{equation}
E_i(p_z) = \left(p_z^{\2} + m_i^{\2}\right)^{\frac{1}{2}}~\nonumber
\end{equation}
and
\begin{equation}
\varphi(p_z - p_i) =  \left(2 \pi a^{\2} \right)^{\frac{1}{4}} \exp{\left[- \frac{(p_z - p_i)^{\2}a^{\2}}{4}\right]}.\nonumber
\end{equation}
In order to obtain the oscillation probability, we must calculate the interference term $\mbox{\sc Int}_{\scalar}(t)$, i. e. we have to
solve the following integral
\begin{eqnarray}
\lefteqn{\mbox{$ \int_{_{\infm}}^{^{\infp}}\frac{dp_z}{2 \pi} \,  \varphi(p_z - p_{\1}) \varphi(p_z - p_{\2})
\exp{[-i \, \Delta E(p_z) \, t]} =$}} \nonumber\\
&\exp{\left[- \frac{(a \, \Delta{p})^{\2}}{8}\right]}\,
\int_{_{\infm}}^{^{\infp}}\frac{dp_z}{2 \pi}  \,
\varphi^{\2}(p_z-\bar{p})\exp{[-i \, \Delta E(p_z) \, t]},~~& \label{6}
\end{eqnarray}
where we have changed the $z$ integration into a $p_z$ integration and introduced the
quantities $\Delta p = p_{\1} - p_{\2}$, $\bar{p} = \frac{1}{2}(p_{\1} + p_{\2})$
and $\Delta E(p_z) = E_{\1}(p_z) - E_{\2}(p_z)$. The oscillation term is
bounded by the exponential function of $a \, \Delta p$ at any
instant of time. Under this condition we could never observe a
{\em pure} flavor-eigenstate. Besides, oscillations are considerably suppressed if $a \, \Delta p > 1$.
A necessary condition to observe oscillations is that
$a \, \Delta p \ll 1$. This constraint can also be expressed by
$\delta p \gg \Delta p$ where $\delta p$ is the momentum
uncertainty of the particle. The overlap between the
momentum distributions is indeed relevant
only for $\delta p \gg \Delta p$. Consequently, without loss of
generality, we can assume
\begin{eqnarray}
\lefteqn{\mbox{\sc Int}_{\scalar}(t) =}\nonumber\\
&~~~~& \mbox{$Re \left[\,\int_{_{\infm}}^{^{\infp}}\frac{dp_z}{2 \pi}\,\varphi^{\2}(p_z - \bar{p})\, \exp{[-i \, \Delta E_i(p_z) \, t]} \, \right]$}
\label{9}.
\end{eqnarray}
In litterature, this equation is often obtained by assuming two
mass eigenstate wave packets described by the ``same'' momentum
distribution centered around the average momentum
$\bar{p}(=p_{\0})$. This simplifying hypothesis  also guarantees
{\em instantaneous} creation of a {\em pure} flavor
eigenstate {\boldmath$\nu_\alpha$} at $t = 0$. In fact, for $
\phi_{\1}(z,0)=\phi_{\2}(z,0)$ we get from Eq.(\ref{0})
\begin{equation}
\phi_{\alpha}(z,0,\theta) = \left(\frac{2}{\pi
a^{\2}}\right)^{\frac{1}{4}} \exp{\left[- \frac{z^{\2}}{a^{\2}}\right]}
\exp{[i \bar{p} \,z]}
\label{9BB}
\end{equation}
and
\begin{equation}\phi_{\beta}(z,0,\theta) =0 \label{9B}.
\end{equation}
To analytically solve the integral in Eq.(\ref{9}), let
us rewrite the energy $E_i(p_z)$ as follows
\begin{eqnarray}
E_i(p_z)  & =  & \mbox{$  \bar{p}\left[1 +
\left(\frac{m_i}{\bar{p}}\right)^{\2} + 2 \left(\frac{p_z -
\bar{p}}{\bar{p}}\right) + \left(\frac{p_z -
\bar{p}}{\bar{p}}\right)^{\2} \right]^{\frac{1}{2}}$} \nonumber \\
 &= & \mbox{$\bar{p}\,  (1 + \chi) \, \left[\, 1 + \frac{\zeta_i}{(1 + \chi)^{\2}}
\, \right]^{\frac{1}{2}}\, \,$} , \label{12}
\end{eqnarray}
where
\begin{equation}
\mbox{$\zeta_i = \left(\frac{m_i}{\bar{p}}\right)^{\2}~~~~
\mbox{and}~~~~ \chi = \frac{p_z - \bar{p}}{\bar{p}}$}.
\label{11}
\end{equation}
In what follows, we shall consider ultra-relativistic particles
and assume a sharply peaked momentum distribution, i. e.
\begin{equation}
m_i \ll \bar{p} ~~\Rightarrow~~\zeta_i \ll 1 ~~~~ \mbox{and}
~~~~~\delta p \ll \bar{p}~~ \Rightarrow ~~\chi \ll 1 \nonumber\\
\label{13}.
\end{equation}
Let us now expand the energy $E_i(p_z)$ in a power series of
$\zeta_i$ and $\chi$. We choose to cut off the power series terms
of order $\zeta_i^{\2}$ $\left(\frac{m_i^{\4}}{\bar{p}^{\4}}\right)$, so that
\begin{eqnarray}
E_i(p_z) &\approx& \mbox{$\bar{p} \left[1 + \chi  + \frac{\zeta_i}{2(1 +
\chi)}\right]$}\nonumber\\
 &=& \bar{p} \left[1 + \chi  + \mbox{$\frac{\zeta_i}{2}$}
 \sum_{j= 0}^{\infty}(-1)^{j}\chi^j\right]. \label{12B}
\end{eqnarray}
In this case, the energy difference becomes
\begin{equation}
\Delta E(p_z)\approx \bar{p} \, \frac{\Delta \zeta}{2}
\sum_{j = 0}^{\infty}(-1)^{j}\chi^j.
\label{140}
\end{equation}
By considering only the first term in the $\chi$ expansion, we reproduce the plane wave result.
Indeed,
\begin{equation}
\Delta E^{ [0]}(p_z)= \bar{p} \, \frac{\Delta \zeta}{2}.
\label{14A}
\end{equation}
An approximation of order $\chi^k$ ($k \geq 1$) in Eq.(\ref{140}) requires some constraints on $\chi$.
Since we cut off terms of order $\Delta \zeta^{\2}$ and we wish to consider terms up to $\chi^k \Delta \zeta$ in Eq.(\ref{140}), it is necessary to satisfy the constraint $\chi^k \Delta \zeta > \frac{\Delta \zeta^{\2}}{2}$ which implies
$\chi > \bar{\zeta}^\frac{1}{k}$ ($\bar{\zeta} = \frac{\zeta_{\1} + \zeta_{\2}}{2}$).
At the same time, for eliminating $\chi^{k+1} \Delta \zeta$, we have to impose
$\chi^{k+1} \Delta \zeta \leq \frac{\Delta \zeta^{\2}}{2}$ which can be rewritten as $\chi \leq \bar{\zeta}^{\frac{1}{1+k}}$.
In this way, an approximation of order $\chi^k$ will be consistent in the range $\bar{\zeta}^{\frac{1}{k}} < \chi \leq \bar{\zeta}^{\frac{1}{k+1}}$.
Meanwhile, the integral in Eq.(\ref{9}) can be solved {\em analytically} only when $k \leq 2$.
By taking into account terms up to the order $\chi^{\2}$, the energy difference becomes
\begin{equation}
\Delta E^{ [2]}(p_z) = \bar{p} \, \frac{\Delta \zeta}{2}\left(1 -\chi + \chi^{\2}\right).
\label{14}
\end{equation}
If we substitute (\ref{14}) in Eq.(\ref{9}) we obtain
\begin{eqnarray}
\lefteqn{\mbox{\sc Int}_{\scalar}(t)
\approx}\nonumber\\
&& \mbox{$Re\left\{\frac{a \bar{p}}{\sqrt{2 \pi}}\int_{_{\infm}}^{^{\infp}} d \chi~\exp{\left[-\frac{(a \bar{p} \chi)^{\2}}{2}\right]}\right.$}\nonumber\\
&& \mbox{$~~~~\left.\times \exp{\left[-i\left(\mathcal{S}(t) + \frac{a \bar{p} \chi}{\sqrt{2}}\mathcal{Q}(t) + \frac{(a \bar{p} \chi)^{\2}}{2}\mathcal{R}(t) \right)\right]}\right\}$}\nonumber\\
&=& \mbox{$Re\left\{\left(\frac{1}{1 + i \mathcal{R}(t)}\right)^{\frac{1}{2}}\exp{\left[-\frac{\mathcal{Q}^{\2}(t)}{4(1 + i \mathcal{R}(t))} - i \mathcal{S}(t)\right]}\right\}$},
\label{15}
\end{eqnarray}
where
\begin{equation}
\mbox{$\mathcal{S}(t) = \frac{\Delta m^{\2}}{2 \bar{p}} t,~~\mathcal{Q}(t) = -\frac{\Delta m^{\2} t}{\sqrt{2} a \bar{p}^{\2}}~~\mbox{and}~~\mathcal{R}(t) = \frac{\Delta m^{\2} t}{a^{\2} \bar{p}^{\3}}$}.
\label{16}
\end{equation}
By suppressing the variable $(t)$ dependence, we can rewrite $\mbox{\sc Int}_{\scalar}(t)$ as
\begin{eqnarray}
\lefteqn{\mbox{\sc Int}_{\scalar}(t)\approx \exp{\mbox{$\left[\min\frac{\mathcal{Q}^{\2}}{4 (1+ \mathcal{R}^{\2})}\right]$}}}\nonumber\\
&&~~~~\times \left\{\mbox{$\sqrt{\frac{(1+ \mathcal{R}^{\2})^\frac{1}{2} + 1}{2(1+ \mathcal{R}^{\2})}}$}\cos{\mbox{$\left[\mathcal{S}\min \frac{\mathcal{Q}^{\2} \mathcal{R}}{4 (1+ \mathcal{R}^{\2})}\right]$}}\min \right.\nonumber\\
&&~~~~~~~~~~~~~~~~~~~~\left. \mbox{$\sqrt{\frac{(1+ \mathcal{R}^{\2})^\frac{1}{2}  \min 1}{2(1+ \mathcal{R}^{\2})}}$}\sin{\mbox{$\left[\mathcal{S} \min \frac{\mathcal{Q}^{\2} \mathcal{R}}{4 (1+ \mathcal{R}^{\2})}\right]$}}\right\}.
\label{19}
\end{eqnarray}

The above result deserve some comments.
The wave packet {\em spreading} is parameterized by $\mathcal{R}(t)$.
At the same time, the {\em slippage} effect between mass eigenstates is predominantly quantified by the $\mathcal{Q}(t)$ parameter.
The {\em spreading} of wave packets is a secondary effect with respect to the {\em slippage} since from Eq.(\ref{16}) we can write
\begin{equation}
\mbox{$\frac{\mathcal{R}(t)}{\mathcal{Q}(t)}\approx \frac{1}{a \bar{p}}$}.
\label{19A}
\end{equation}
Under {\em minimal spreading} conditions, i. e. when $\mathcal{R} \ll 1$, the Eq.(\ref{19}) becomes
\begin{eqnarray}
\lefteqn{\mbox{\sc Int}_{\scalar}(t)  \approx
\mbox{$\exp{\left[\min\frac{\mathcal{Q}^{\2}\left(1 - \mathcal{R}^{\2}\right)}{4}\right]}$}}\nonumber\\
&&\times\left\{\mbox{$\left(1 \min \frac{3 \mathcal{R}^{\2}}{8}\right)$}\cos{\mbox{$\left[\mathcal{S} \min \frac{\mathcal{Q}^{\2} \mathcal{R}}{4}\right]$}} \min \mbox{$\frac{\mathcal{R}}{2}$}\sin{\mbox{$\left[\mathcal{S} \min \frac{\mathcal{Q}^{\2} \mathcal{R}}{4}\right]$}}\right\},
\label{19C}
\end{eqnarray}
where the oscillating character is predominantly given by the cosine function behavior.
The exponential term with $\mathcal{R}(t)$ extends the interference between the mass eigenstate wave packets, and consequently the oscillating character, for (a little) longer times.
Taking into account terms up to the order $\chi$ in the Eq.(\ref{140}), we can write
\begin{equation}
\Delta E^{ [1]}(p_z)= \bar{p} \, \frac{\Delta \zeta}{2}(1 - \chi)
\label{14AA}
\end{equation}
and compute the oscillation probability with the leading corrections due to the {\em slippage} effect,
\begin{eqnarray}
\lefteqn{P_{\scalar}(\mbox{\boldmath$\nu_\alpha$}\rightarrow\mbox{\boldmath$\nu_\beta$};t)
 \approx}\nonumber\\
 &&~~~~~~~~~~~~~~ \mbox{$\frac{\sin^{\2}{[2\theta]}}{2}\left\{1- \exp{\left[-\frac{\mathcal{Q}^{\2}(t)}{4}\right]}\cos{[\mathcal{S}(t)]}\right\}$},
\label{20}
\end{eqnarray}
which corresponds to the same result obtained by \cite{DeL04}.
Under {\em minimal slippage} conditions, i. e. when $\mathcal{Q}(t) \ll 1$, the Eq.(\ref{20}) reproduces the plane wave formula (\ref{0000}).
\begin{eqnarray}
P_{\scalar}(\mbox{\boldmath$\nu_\alpha$}\rightarrow\mbox{\boldmath$\nu_\beta$};t)
 &\approx& \mbox{$\frac{\sin^{\2}{[2\theta]}}{2}\left\{1- \left(1 -\frac{\mathcal{Q}^{\2}(t)}{4}\right)\cos{\left[\mathcal{S} t \right]}\right\}$}\nonumber\\
 &\approx&  \mbox{$\frac{\sin^{\2}{[2\theta]}}{2}\left\{1- \cos{[\mathcal{S}(t)]}\right\}$}\nonumber\\
 &=& \mbox{$\sin^{\2} [2 \theta] \sin^{\2} \left[ \frac{\Delta m^{\2}}{4 \bar{p}} \, t \right]$}.
\label{20A}
\end{eqnarray}


\section{Dirac formalism}

The results in the previous section have been obtained by considering {\em scalar} mass eigenstates.
Neutrinos are, however, {\em fermions}.
The time evolution of a spin one-half particle have to be described by the Dirac equation.
To introduce the {\em fermionic} character in the study of quantum oscillation phenomena, we shall use the Dirac equation as the evolution equation for the mass eigenstates.
The Eq.(\ref{0}) now becomes
\begin{eqnarray}
\Psi(z,t) &=& \psi_{\1}(z,t)\cos{\theta}\,\mbox{\boldmath$\nu_{\1}$} + \psi_{\2}(z,t)\sin{\theta}\,\mbox{\boldmath$\nu_{\2}$}\nonumber\\
          &=& \left[\psi_{\1}(z,t)\cos^{\2}{\theta} + \psi_{\2}(z,t)\sin^{\2}{\theta}\right]\,\mbox{\boldmath$\nu_\alpha$} \nonumber\\
          && ~~~~~~~~~~~~+ \left[\psi_{\1}(z,t) - \psi_{\2}(z,t)\right]\cos{\theta}\sin{\theta}\,\mbox{\boldmath$\nu_\beta$}\nonumber\\
          &=& \psi_{\alpha}(z,t;\theta)\,\mbox{\boldmath$\nu_\alpha$} + \psi_{\beta}(z,t;\theta)\,\mbox{\boldmath$\nu_\beta$},
\label{0B}
\end{eqnarray}
where $\psi_i(z,t)$ satisfies the Dirac equation for a mass $m_i$.
The natural extension of Eq.(\ref{9B}) reads
\begin{equation}
\psi_{\alpha}(z,0,\theta) = \phi_{\alpha}(z,0,\theta) w
\label{22}
\end{equation}
where $w$ is a constant spinor which satisfies the normalization condition $w^{\dagger} w = 1$.

\subsection{Dirac wave packets and the oscillation formula}

To describe the time evolution of mass eigenstate Dirac wave packets,
we could be inclined to superpose only positive frequency solutions of the Dirac equation.
It seems, at first glance, a reasonable choice.
However, when the initial state has the form given in Eq.(\ref{22}), it is necessary to
superpose both positive and negative frequency solutions of Dirac equation.
Let us clear up this point.
The flavor state $\psi_{\alpha}(z,t,\theta)$ is now expressed in terms of
\begin{eqnarray}
\psi_i(z,t)
            &=& \mbox{$\int_{_{\infm}}^{^{\infp}}\frac{dp_z}{2\pi}$} \exp{[ i p_z z]}\nonumber\\
            &&~~ \times\sum_{s=1,2}\{b^s_i(p_z)u^s_i(p_z) \exp{[-iE_i(p_z) t]}\nonumber\\
            &&~~~~~~~+ d^{s*}_i(\mi p_z)v^s_i(\mi p_z) \exp{[+iE_i(p_z)t]}\}.~~~~
\label{23}
\end{eqnarray}
At time $t=0$ the mass eigenstate wave functions satisfy $\psi_{\1}(z,0)=\psi_{\2}(z,0)$ (this
guarantees the {\em instantaneous} creation of a {\em pure} flavor-eigenstate {\boldmath$\nu_\alpha$} as we have appointed in section II).
The Fourier transform of $\psi_i(z,0)$ is
\begin{eqnarray}
\sum_{s=1,2}\left[b^{s}_i(p_z)u^{s}_i(p_z) + d^{s*}_i(\mi p_z)v^{s}_i(\mi p_z)\right].
\label{24}
\end{eqnarray}
By observing that the Fourier transform of $\phi_{\alpha}(z,0,\theta)$ is given by $\varphi(p_z - \bar{p})$ (see Eq.(\ref{9B})), we immediately obtain the Fourier transform of $\psi_{\alpha}(z,0,\theta)$,
\begin{eqnarray}
\varphi(p_z - \bar{p}) w & = &\sum_{s=1,2}\left[b^{s}_i(p_z)u^{s}_i(p_z) + d^{s*}_i(\mi p_z)v^{s}_i(\mi p_z)\right].
\label{25}
\end{eqnarray}
Using the orthogonality properties of Dirac spinors, we find \cite{Zub80}
\begin{eqnarray}
b^s_i(p_z) &=& \varphi(p_z - \bar{p})u^{s \dagger}_i(p_z) w, \nonumber\\
d^{s*}_i(\mi p_z) &=& \varphi(p_z - \bar{p})v^{s \dagger}_i(\mi p_z) w.
\label{26}
\end{eqnarray}
These coefficients carry an important physical information.
For {\em any} initial state which has the form given in Eq.(\ref{22}), the the negative frequency solution coefficients $d^{s*}_i(\mi p_z)$ necessarily provides a non-null contribution to the time evolving wave packet.
This obliges us to take the complete set of Dirac equation solutions to construct the wave packet.
Only if we consider a momentum distribution given by a delta function (plane wave limit) and suppose an initial spinor $w$ being a positive energy mass eigenstate with momentum $\bar{p}$, the contribution due to $d^{s*}_i(\mi p_z)$ will be null.

Having introduced the Dirac wave packet prescription, we are now in a position to calculate the flavor conversion formula.
The following calculations do not depend on the gamma matrix representation.
By substituting the coefficients given by Eq.(\ref{26}) in Eq.(\ref{23}) and using the well-known spinor properties \cite{Zub80},
\begin{eqnarray}
&&\sum_{s=1,2}u^s_i(p_z)\overline{u}^s_i(p_z) = \frac{\gamma^0 E_i(p_z) - \gamma^{\3} p_z + m_i}{2E_i(p_z)}, \nonumber\\
&&\sum_{s=1,2}v^s_i(\mi p_z)\overline{v}^s_i(\mi p_z) = \frac{\gamma^0 E_i(p_z) + \gamma^{\3} p_z -m_i}{2E_i(p_z)},
\label{28}
\end{eqnarray}
we obtain
\begin{eqnarray}
\psi_i(z,t) & = &\int_{_{\infm}}^{^{\infp}}\frac{dp_z}{2 \pi} \, \varphi(p_z -\bar{p}) \exp{[ip_z z]}
\{\cos{[E_i(p_z) t]} \nonumber\\
&&~~~~~~~~  -\frac{i\gamma^0\left(\gamma^{\3} p_z+ m_i\right)}{E_i(p_z)}\sin{[E_i(p_z) t]}\} w.
\label{29}
\end{eqnarray}
By simple mathematical manipulations, the new interference oscillating term will be written as
\begin{eqnarray}
\mbox{\sc Int}_{\Dirac}(t) & = & \mbox{$\int_{_{\infm}}^{^{\infp}}\frac{dp_z}{2 \pi} \, \varphi^{\2}(p_z -\bar{p})$}\nonumber\\
&&~~\times\left\{\left(1 -  F(p_z)\right) \cos{[\Delta E(p_z) t]} \right.\nonumber\\
&& \left.~~~~~~~~~~~~~+ F(p_z) \cos{[2\bar{E}(p_z) t]}\right\}
\label{30}
\end{eqnarray}
where
\begin{equation}
\bar{E}(p_z) = \frac{E_{\1}(p_z) + E_{\2}(p_z)}{2}\nonumber
\end{equation}
and
\begin{equation}
F(p_z) = \frac{1}{2} - \frac{ p_z^{\2} + m_{\1}m_{\2}}{2 E_{\1}(p_z)E_{\2}(p_z)}.\nonumber
\end{equation}
What is interesting about the result in Eq.(\ref{30}) is that it was obtained without any assumption on the initial spinor $w$.
Otherwise, the initial spinor carries some fundamental physical information about the created state.
And this could be relevant in the study of chiral oscillations \cite{DeL98} where the initial state plays a fundamental role.
With respect to the standard treatment of neutrino oscillations done by using
{\em scalar} wave packets and leading to the interference term
\begin{equation}
\mbox{\sc Int}_{\scalar}(t) = \int_{_{\infm}}^{^{\infp}}\frac{dp_z}{2 \pi}
 \, \varphi^{\2}(p_z - \bar{p})\cos{[ \Delta E(p_z) \, t]},
\label{34A}
\end{equation}
we note in $\mbox{\sc Int}_{\Dirac}(t)$ two additional terms.
In the first one, the {\em standard} oscillating term $\cos{[\Delta E(p_z)\, t]}$,
which arises from the interference between mass eigenstate components of
equal sign frequencies, is multiplied by a {\em new factor} obtained by
the products
\begin{eqnarray}
u_{\1}^{\dagger}(p_z)\,u_{\2}(p_z), ~~v_{\1}^{\dagger}(\mi p_z)\,v_{\2}(\mi p_z)~~\mbox{and h.c.}.\nonumber
\end{eqnarray}
The second one, $\cos{[2 \bar{E}(p_z) t]}$, is a {\em new} oscillating term
which comes from the interference between mass eigenstate components of positive and negative frequencies.
The factor multiplying such an additional oscillating term is obtained by the products
\begin{eqnarray}
u_{\1}^{\dagger}(p_z)\,v_{\2}(\mi p_z), ~~v_{\1}^{\dagger}(\mi p_z)\,u_{\2}(p_z)~~\mbox{and h.c.}.\nonumber
\end{eqnarray}
The new oscillations have very high frequencies.
Such a peculiar oscillating behavior is similar to the phenomenon referred to as {\em Zitterbewegung}.
In atomic physics, the electron exhibits this violent quantum fluctuation in the position and
becomes sensitive to an effective potential which explains the Darwin term in the hydrogen atom \cite{Sak87}.
We shall see later that, at the instant of creation, such rapid oscillations introduce a small
modification in the oscillation formula.

\begin{figure}
\resizebox{0.47\textwidth}{!}{%
  \includegraphics{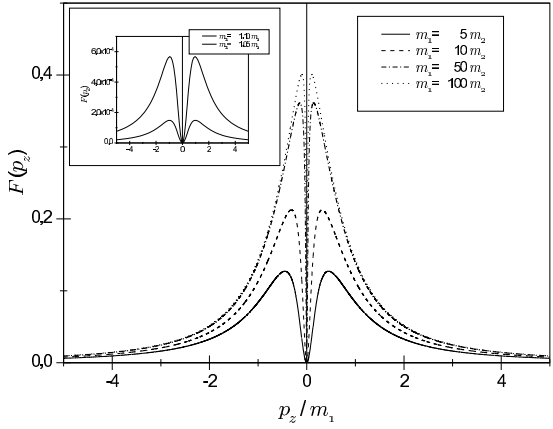}
}
\caption{The
function $F(p_z)$ is plotted for different values of the ratio
between $m_{\1}$ and $m_{\2}$. For a momentum distribution sharply
peaked around $\bar{p} \gg m_{\1,\2}$, $F(p_z)$ does not play a
significant role in the ``modified'' oscillation formula. In the
case of $m_{\1} \approx m_{\2}$, independently of the value of $\bar{p}$
and of the momentum distribution width, the maximum values of
$F(p_z)$ are  negligible and consequently $F(p_z)$ is practically
suppressed in the calculation (see amplification in the upper
box).}
\label{fig1}       
\end{figure}

We plot the function $F(p_z)$ in Figure \ref{fig1}.
We can readily observe that it goes rapidly to
zero for $p_z \gg m_{\1,\2}$, it has a minimum  at $p_z=0$ and
two maxima at $p_z = \pm\sqrt{m_{\1} m_{\2}}$.
The maximum value of $F(p_z)$ is
\begin{equation}
F_{max}(p_z) = \frac{1}{2}\left(1 - \frac{\sqrt{m_{\1}m_{\2}}}{m_{\1} +m_{\2}}\right)
\label{36}
\end{equation}
which vanishes in the limit $m_{\1} = m_{\2}$.
As we can see in Figure \ref{fig1}, the new effects are relevant only when $\Delta m \approx m_{\1} \gg m_{\2}$.

\subsection{The oscillation formula with \textit{spreading}}

To quantify the new effects exhibited in the oscillation probability formula let us calculate the integral of Eq.(\ref{30}) by assuming an ultra-relativistic particle and following the localization condition given by (\ref{13}).
We use the same criteria adopted in section II to expand the energy (\ref{12}) in a power series of $\zeta_i$ and $\chi$.
A second order approximation in $\chi$ allows us to write the energy dependent terms as
\begin{eqnarray}
\label{A1}
\Delta E^{ [2]}(p_z)&=&\mbox{$ \frac{\Delta m^{\2}}{2\bar{p}}(1 - \chi + \chi^{\2})$},\\
\label{A2}
\bar{E}^{ [2]}(p_z) &=& \mbox{$\bar{p}\left[\left(2 + \frac{m_{\1}^{\2} + m_{\2}^{\2}}{2 \bar{p}^{\2}}\right) \right.$}\nonumber\\
&&\mbox{$~~~~~~\left. + \left(2 -\frac{m_{\1}^{\2} +m_{\2}^{\2}}{2 \bar{p}^{\2}}\right)\chi + \frac{m_{\1}^{\2} +m_{\2}^{\2}}{2 \bar{p}^{\2}}\chi^{\2}\right]$},\\
\label{A3}
F^{ [2]}(p_z)       &=& \mbox{$\left(\frac{\Delta m}{2\bar{p}}\right)^{\2} (1 - 2 \chi + 3\chi^{\2})$}.
\end{eqnarray}
Since we have approximated not only $\Delta E(p_z)$, but also $F(p_z)$ and $E(p_z)$, the range of validity for an analytical approximation of order $\chi^k$ is now given by $\left(\frac{\bar{\zeta}^{\2}}{\Delta \zeta}\right)^{\frac{1}{k}} < \chi \leq \bar{\zeta}^{\frac{1}{k+1}}$.
By substituting $\varphi(p_z -\bar{p})$ and the approximations (\ref{A1}-\ref{A3}) in Eq.(\ref{30}), we obtain
\begin{eqnarray}
\lefteqn{\mbox{\sc Int}_{\Dirac}(t) \approx \mbox{$\frac{a \bar{p}}{(2\pi)^{\frac{1}{2}}}\int_{_{\infm}}^{^{\infp}} d\chi\,
 \exp{\left[-\frac{(a \bar{p} \chi)^{\2}}{2}\right]}$}}\nonumber\\
&&~\times \left\{
\mbox{$\left[1 - \left(\frac{\Delta m}{2\bar{p}}\right)^{\2}(1 -2 \chi + 3\chi^{\2})\right]$}\right.\nonumber\\
&&~~~\times\cos{\mbox{$\left[\frac{\Delta m^{\2}\,t}{2\bar{p}} (1 - \chi + \chi^{\2})\right]$}} \nonumber\\
&& ~ +  \mbox{$\left(\frac{\Delta m}{2\bar{p}}\right)^{\2} (1 -2 \chi + 3\chi^{\2})$}\nonumber\\
&&\left. ~~~ \times\cos{\mbox{$\left[ \bar{p} t \left( 2 ( 1 + \chi) + \frac{m_{\1}^{\2} +m_{\2}^{\2}}{2 \bar{p}^{\2}} (1 - \chi + \chi^{\2})\right)\right]$}}\right\}.~
\label{A4}
\end{eqnarray}
A new integrating variable $\sigma = \frac{a \bar{p} \chi}{\sqrt{2}}$ and the coefficients
\begin{equation}
\mbox{$
\begin{array}{lll}
\mathcal{S}_{\mi}(t) = \frac{\Delta m^{\2}}{2\bar{p}} t,&
&\mathcal{S}_{\pl}(t) = \bar{p} t \left(2 + \frac{m_{\1}^{\2} +m_{\2}^{\2}}{2 \bar{p}^{\2}}\right),\\
\mathcal{Q}_{\mi}(t) = -\frac{\Delta m^{\2}}{\sqrt{2}a\bar{p}^{\2}} t,&
&\mathcal{Q}_{\pl}(t) =
\frac{\sqrt{2}\bar{p} t}{a \bar{p}} \left(2 - \frac{m_{\1}^{\2} +m_{\2}^{\2}}{2 \bar{p}^{\2}}\right),\\
\mathcal{R}_{\mi}(t) =\frac{\Delta m^{\2}}{a^{\2}\bar{p}^{\3}} t,&
\mbox{~~~~}&\mathcal{R}_{\pl}(t) = \bar{p} t \left(\frac{m_{\1}^{\2} +m_{\2}^{\2}}{a^{\2} \bar{p}^{\4}}\right)
\end{array}$}
\label{A6}
\end{equation}
enable us to write Eq.(\ref{A4}) in the form
\begin{eqnarray}
\lefteqn{\mbox{\sc Int}_{\Dirac}(t) \approx \mbox{$ \int_{_{\infm}}^{^{\infp}}\frac{d \sigma}{\sqrt{\pi}}\, \exp{\left[- \sigma^{\2}\right]}$}}\nonumber\\
&&\times Re\left\{\mbox{$\left[1 - \left(\frac{\Delta m}{2\bar{p}}\right)^{\2}\left(1 - \frac{2\sqrt{2}}{a\bar{p}}\sigma + \frac{6}{(a\bar{p})^{\2}}\sigma^{\2} \right)\right]$}\right.\nonumber\\
&&~~~~~~\times\exp{\left[-i\mathcal{S}_{\mi}(t)  -i\mathcal{Q}_{\mi}(t)\sigma  -i\mathcal{R}_{\mi}(t)\sigma^{\2}\right]}\nonumber\\
&& ~~ + \mbox{$\left(\frac{\Delta m}{2\bar{p}}\right)^{\2} \left(1 - \frac{2\sqrt{2}}{a\bar{p}}\sigma + \frac{6}{(a\bar{p})^{\2}}\sigma^{\2} \right)$}\nonumber\\
&&\left.~~~~~~ \times\exp{\left[-i\mathcal{S}_{\pl}(t)  -i\mathcal{Q}_{\pl}(t)\sigma  -i\mathcal{R}_{\pl}(t)\sigma^{\2}\right]}\right\}\nonumber\\
&&~~~~~~~~~~~~~= Re\left[H_{\mi}(t)G_{\mi}(t) + H_{\pl}(t)G_{\pl}(t)\right]
\label{A7}
\end{eqnarray}
where
\begin{eqnarray}
G_{\ppm}(t) &=& \mbox{$\left(\frac{1}{1 + i \mathcal{R}_{\ppm}(t)}\right)^{\frac{1}{2}}\exp{\left[-\frac{\mathcal{Q}^{\2}_{\ppm}(t)}{4(1 + i \mathcal{R}_{\ppm}(t))} - i \mathcal{S}_{\ppm}(t)\right]}~~$}
\label{A9}
\end{eqnarray}
are obtained in the same way as (\ref{20}), and
\begin{eqnarray}
H_{\ppm}(t) &=& \mbox{$\frac{1}{2} \mp \left\{\frac{1}{2} - \left(\frac{\Delta m}{2\bar{p}}\right)^{\2}
\left[1  + i \frac{\sqrt{2}}{a \bar{p}}\frac{\mathcal{Q}_{\ppm}(t)}{1 + i \mathcal{R}_{\ppm}(t)}\right.\right.$} \nonumber\\
&&\mbox{$\left.\left. ~~+ \frac{3}{(a\bar{p})^{\2}}\left(\frac{1}{1 + i \mathcal{R}_{\ppm}(t)} -
  \frac{\mathcal{Q}^{\2}_{\ppm}(t)}{2(1 + i \mathcal{R}_{\ppm}(t))^{\2}}\right)\right]\right\}$}\nonumber\\
\label{A10}
\end{eqnarray}
arise from the new coefficients which include $F(p_z)$.

\subsection{The oscillation formula without \textit{spreading}}

A more satisfactory interpretation of the modifications introduced by the Dirac formalism is given when we restrict our study to a first order approximation in $\chi$, i. e. without considering the wave packet {\em spreading}.
In fact, we could take into account terms up to the order $\chi$ in the Eqs.(\ref{A1}-\ref{A3}) and obtain a simpler approximation,
\begin{eqnarray}
\mbox{\sc Int}_{\Dirac}(t)
&\approx&\mbox{$ \int_{_{\infm}}^{^{\infp}}\frac{d \sigma}{\sqrt{\pi}}\,
Re \left\{\left[1 - \left(\frac{\Delta m}{2\bar{p}}\right)^{\2}\left(1 - \frac{2\sqrt{2}}{a\bar{p}}\sigma \right)\right]\right.$}\nonumber\\
&&\mbox{$~~~~~~~~~~\times \exp{\left[-i\mathcal{S}_{\mi}(t) - i \mathcal{Q}_{\mi}(t)\sigma - \sigma^{\2}\right]}$} \nonumber\\
&&\mbox{$~~~~ + \left(\frac{\Delta m}{2\bar{p}}\right)^{\2}\left(1 - \frac{2\sqrt{2}}{a\bar{p}}\sigma \right)$}\nonumber\\
&& \mbox{$\left.~~~~~~~~~~\times\exp{\left[-i\mathcal{S}_{\pl}(t)  - i \mathcal{Q}_{\pl}(t)\sigma - \sigma^{\2}\right]}\right\}$}\nonumber\\
&=&\mbox{$ Re\left\{\left[1 - \left(\frac{\Delta m}{2\bar{p}}\right)^{\2}\left(1  + i \frac{\sqrt{2}}{a \bar{p}}\mathcal{Q}_{\mi}(t)\right)\right]\right.$}\nonumber\\
&&\mbox{$~~~~~~~~~~~~\times\exp{\left[-\frac{\mathcal{Q}^{\2}_{\mi}(t)}{4} - i \mathcal{S}_{\mi}(t)\right]}$}\nonumber\\
& &\mbox{$~~~~+ \left(\frac{\Delta m}{2\bar{p}}\right)^{\2}\left(1  + i \frac{\sqrt{2}}{a \bar{p}}\mathcal{Q}_{\pl}(t)\right)$}\nonumber\\
&& \mbox{$\left.~~~~~~~~~~~~\times\exp{\left[-\frac{\mathcal{Q}^{\2}_{\pl}(t)}{4} - i \mathcal{S}_{\pl}(t)\right]}\right\},$}
\label{A11}
\end{eqnarray}
By using the explicit expressions for $\mathcal{Q}_{\ppm}(t)$ and $\mathcal{S}_{\ppm}(t)$ we get
\begin{eqnarray}
\lefteqn{\mbox{\sc Int}_{\Dirac}(t) \approx }\nonumber\\
& &\mbox{$\exp{\left[-\left(\frac{\Delta m^{\2}\, t}{2 \sqrt{2}a \bar{p}^{\2}}\right)^{\2}\right]}$}
\mbox{$\left\{\left[1 - \left(\frac{\Delta m}{2\bar{p}}\right)^{\2}\right]\cos{\left[\frac{\Delta m^{\2}}{2 \bar{p}}t\right]}\right.$}\nonumber\\
&&\mbox{$\left.~~~~~~~~~~~~~~~~~~~~~~~~~~ + \left(\frac{\Delta m}{2\bar{p}}\right)^{\2} \frac{\Delta m^{\2}}{a^{\2}\bar{p}^{\3}}t\sin{\left[\frac{\Delta m^{\2}}{2 \bar{p}}t\right]}\right\}$}\nonumber\\
&+&\mbox{$\exp{\left[-\frac{t^{\2}}{2 a^{\2}}\left(2 - \frac{m_{\1}^{\2} + m_{\2}^{\2}}{2 \bar{p}^{\2}}\right)^{\2}\right]}
~$}\nonumber\\
&&\mbox{$~\times \left(\frac{\Delta m}{2\bar{p}}\right)^{\2}
\left\{\cos{\left[\bar{p}t\left(2 + \frac{m_{\1}^{\2} + m_{\2}^{\2}}{2 \bar{p}^{\2}}\right)\right]}\right.$}\nonumber\\
&&\mbox{$\left.~~ + \frac{2\bar{p}t}{(a \bar{p})^{\2}}\left(2 - \frac{m_{\1}^{\2} + m_{\2}^{\2}}{2 \bar{p}^{\2}}\right)\sin{\left[\bar{p}t\left(2 + \frac{m_{\1}^{\2} + m_{\2}^{\2}}{2 \bar{p}^{\2}}\right)\right]}\right\}$}.
\label{A14}
\end{eqnarray}
As we have already noticed, the oscillating functions going with the second exponential function in Eq.(\ref{A14}) arise from the interference between positive and negative frequency solutions of the Dirac equation.
It produces very high frequency oscillations which is similar to the quoted phenomenon of {\em Zitterbewegung} \cite{Sak87}.
The oscillation length which characterizes the very high frequency oscillations is given by $L^{ VHF}_{0sc} \approx \frac{2 \pi}{\bar{p}}$.
Obviously, $L^{ VHF}_{0sc}$ is much smaller than the standard oscillation length given by $L^{Std}_{0sc} = \frac{4 \pi \bar{p}}{\Delta m^{\2}}$.
It means that the propagating particle exhibits a violent quantum fluctuation of its flavor quantum number around a flavor average value which oscillates with $L^{Std}_{0sc}$.
Meanwhile, except at times $t \sim 0$, it provides a practically null contribution to the oscillation probability.
To explain such a statement, let us suppose that an experimental measurement takes place after a time $T \approx L$ for ultra-relativistic particles.
The observability conditions impose that the propagation distance $L$ must be larger than the wave packet localization $a$.
Since the (second) exponential function vanishes when $L \gg a$, for measurable distances,
the effective flavor conversion formula will not contain such very high frequency oscillation terms, and can be written as
\begin{eqnarray}
\lefteqn{P_{\Dirac}(\mbox{\boldmath$\nu_\alpha$}\rightarrow\mbox{\boldmath$\nu_\beta$};L) \approx\mbox{$\frac{\sin^{\2}{[2\theta]}}{2}$}} \nonumber\\
&& \mbox{$\times\left\{ 1 -\exp{\left[-\left(\frac{\Delta m^{\2}\, L}{2 \sqrt{2}a \bar{p}^{\2}}\right)^{\2}\right]}
 \left\{\left[1 - \left(\frac{\Delta m}{2\bar{p}}\right)^{\2}\right]\cos{\left[\frac{\Delta m^{\2}}{2 \bar{p}}L\right]}~~ \right.\right.$}\nonumber\\
&& \mbox{$\left.\left. ~~~~~~~~~~~~~~~~~~~~~~ + \left(\frac{\Delta m}{2\bar{p}}\right)^{\2} \frac{\Delta m^{\2}}{a^{\2}\bar{p}^{\3}}L\sin{\left[\frac{\Delta m^{\2}}{2 \bar{p}}L\right]}\right\}\right\}$}.
\label{A14BB}
\end{eqnarray}
For distances which are restrict to the interval $a \ll L \ll a \frac{2 \sqrt{2} \bar{p}^{\2}}{\Delta m^{\2}}$ we observe the {\em minimal slippage} between the wave packets.
In this case, we could suddenly approximate the oscillation probability to
\begin{eqnarray}
\lefteqn{P_{\Dirac}(\mbox{\boldmath$\nu_\alpha$}\rightarrow\mbox{\boldmath$\nu_\beta$};L)
 \approx }\nonumber\\
&& \mbox{$\frac{\sin^{\2}{[2\theta]}}{2}\left\{
1 - \left[1-\left(\frac{\Delta m^{\2} L}{2 \sqrt{2}a \bar{p}^{\2}}\right)^{\2}\right] \right.$}\nonumber\\
&&\mbox{$\left.~~~~~~~~~~~~~~~~~~~~~~\times\left[1 - \left(\frac{\Delta m}{2\bar{p}}\right)^{\2}\right]\cos{\left[\frac{\Delta m^{\2}}{2 \bar{p}}L\right]} \right\}$},
\label{A14B}
\end{eqnarray}
however, we reemphasize that it is {\em not} valid for $T \approx L \sim 0$ when the rapid oscillations are still relevant ($L < a$).
By comparing the result of Eq.(\ref{A14B}) with the {\em scalar} oscillation probability of Eq.(\ref{20}),
we notice a deviation of the order $\left(\frac{\Delta m}{2\bar{p}}\right)^{\2}$ that appears as an additional coefficient of the cosine function.
It is not relevant in the ultra-relativistic limit as we have noticed after studying the function $F(p_z)$.

\subsection{A brief extension to quantum field treatment}

To finalize our study, we try to establish a tenuous
correspondence between our results and the QFT treatment.
It was extensively demonstrated in the literature \cite{Ric93,Giu93,Giu02B}
that the oscillating particle cannot be treated in isolation.
The oscillation process must be considered globally: the oscillating
states become intermediate states, not directly observed,
which propagate between a {\em source}
and a {\em detector}. This idea can be implemented in QFT
when the intermediate
oscillating states are represented by internal lines of
Feynman diagrams
and the interacting particles at source/detector are described
by {\em external} wave packets \cite{Giu93,Beu03}.
In this context, let us consider the weak flavor-changing processes
occurring through the intermediate propagation of a neutrino,
\begin{equation}
P_I \rightarrow P_F + \alpha + \nu_{\alpha}
 ~~(oscillation)~~
 \nu_{\beta} + D_I \rightarrow \beta + D_F
\label{A15}
\end{equation}
where $P_I$ and $P_F$ ($D_I$ and $D_F$) are respectively
the initial and final production (detection) particles.
The amplitude for the process is represented by
\begin{equation}
\mathcal{A} = \mbox{$\left\langle P_F, D_F \left|\mathbf{T}\left(
\exp{\left[-i\,\int{dx^{\4} \, \mathcal{H}_I}\right]}\right)
- \mathbf{1}\right| P_I, D_I \right\rangle$}
\label{A16}
\end{equation}
where $\mathcal{H}_I$ is the interaction Hamiltonian for the
intermediate particle and $\mathbf{T}$ is the time ordering
operator.
After some mathematical manipulations \cite{Beu03},
this amplitude can be represented by the integral
\begin{eqnarray}
\mathcal{A} &=& \int{\frac{dE\, d\mathbf{p}^{\3}}{(2\pi)^{\4}}\,
F(E,\mathbf{p})}\nonumber\\
&&~~~~~\times G(E,\mathbf{p},t_D,t_P)\,
\exp{[i\, \mathbf{p}\cdot(\mathbf{x}_D - \mathbf{x}_P)]}
\label{A17}
\end{eqnarray}
where the function $F(E,\mathbf{p})$ represents the {\em overlap}
of the incoming and outgoing wave
packets, both at the source and at the detector, and
the {\em Green} function in the momentum space,
$G(E, \mathbf{p}, t_D, t_P)$, represents the fermion propagator which carries the
information of the oscillation process.
The overlap function is independent of production
and detection times and positions
($t_P$, $t_D$, $\mathbf{x}_P$, $\mathbf{x}_D)$ and depends on the
the directions of incoming and outgoing momenta.
In certain way, the physical conditions of source and detector,
in terms of time and space intervals, are better defined in this framework
than in the {\em intermediate} wave packet framework.
Anyway, to understand the oscillation process
we must turn back to the definition of mixing in quantum mechanics.
It is similar in field theory, except
that it applies to fields, not to physical states.
This difference allows to bypass the problems
arising in the definition of flavor and mass bases \cite{Beu03}.
In one-dimensional spatial coordinates, the mixing
is illustrated by the unitary transformation
\begin{equation}
\psi_{\sigma}(z,t;\theta) = \mathcal{G}^{\mi \1}(\theta; t)\,
\psi_i(z,t)\,\mathcal{G}(\theta; t)
\label{A00}
\end{equation}
as the result of the noncoincidence of the flavor basis
($\sigma =\,\alpha, \, \beta$)
and the mass basis ($i =\, 1,\, 2$).
The Eq.(\ref{A00}) gives the the Eq.(\ref{0B})
when the generator of mixing transformations
$\mathcal{G}(\theta; t)$ is given by
\begin{eqnarray}
\mathcal{G}(\theta; t) &=& \exp[\theta \int\,dz \, \psi_{\1}(z,t)\psi_{\2}(z,t)\nonumber\\
&&~~~~~~~~~~~~~~~~~~~~~~-\psi_{\2}(z,t)\psi_{\1}(z,t)]
\label{A00B}.
\end{eqnarray}
By taking the one-dimensional representation of Eq.(\ref{A17}),
the propagator $G(E,p_z,t_D,t_P)$
can also be written in the flavor basis as
\begin{eqnarray}
G^{\alpha\beta}(\theta; E,p_z,T) &=&
 \mathcal{G}^{\mi \1}(\theta; t)\,G(E,p_z,T)\,\mathcal{G}(\theta; t)\nonumber\\
&=& \mathcal{G}^{\mi \1}(\theta; t)\,G(E,p_z,t_D,t_P)\,\mathcal{G}(\theta; t)~~
\label{A00C}
\end{eqnarray}
with $T = t_D - t_P$.

In particular, by following the Blasone and Vitiello (BV)
prescription \cite{Bla95,Bla03B},
the definition of a Fock space of weak eigenstates becomes possible
and a nonperturbative flavor oscillation amplitude can be derived.
In this case, the complete Lagrangian (density) is split in
a propagation Lagrangian,
\begin{eqnarray}
\mathcal{L}_{p}(z,t) &=&
\bar{\psi}_{\1}(z,t)\,\left(i \,\partial\hspace{-0.2cm}\slash\hspace{0.1cm} - m_{\1}\right)\,\psi_{\1}(z,t)\nonumber\\
&&~~~~+\bar{\psi}_{\2}(z,t)\,\left(i \,\partial\hspace{-0.2cm}\slash\hspace{0.1cm} - m_{\2}\right)\,\psi_{\2}(z,t),
\label{A21}
\end{eqnarray}
and an interaction Lagrangian
\begin{eqnarray}
\mathcal{L}_{i}(z,t)&=&
\bar{\psi}_{\alpha}(z,t;\theta)\,\left(i \,\partial\hspace{-0.2cm}\slash\hspace{0.1cm} - m_{\alpha}\right)\,\psi_{\alpha}(z,t;\theta)\nonumber\\
&&~~~~+\bar{\psi}_{\beta}(z,t;\theta)\,\left(i \,\partial\hspace{-0.2cm}\slash\hspace{0.1cm} - m_{\beta}\right)\,\psi_{\beta}(z,t;\theta)\nonumber\\
&&~~~~- m_{\alpha\beta} \,\left(\bar{\psi}_{\alpha}(z,t;\theta)\psi_{\beta}(z,t;\theta) \right.\nonumber\\
&&\left.~~~~~~~~~~~~~~~~~~+ \bar{\psi}_{\beta}(z,t;\theta)\psi_{\alpha}(z,t;\theta)\right),
\label{A20}
\end{eqnarray}
where
\begin{eqnarray}
m_{\alpha (\beta)} = m_{\1(\2)}\, \cos^{\2}{\theta} + m_{\2(\1)}\,\sin^{\2}{\theta}\nonumber
\end{eqnarray}
and\begin{eqnarray}
m_{\alpha\beta} = (m_{\1} - m_{\2})\, \cos{\theta}\sin{\theta}.\nonumber
\end{eqnarray}
In general, the two subsets of the Lagrangian can be distinguished if there is a flavor
transformation which is a symmetry of $\mathcal{L}_{i}(z,t)$ but not of $\mathcal{L}_{p}(z,t)$.
Particle mixing occurs if the propagator built from $\mathcal{L}_{p}(z,t)$,
and representing the creation of a particle of flavor $\alpha$ at point
$z$ and the annihilation of a particle of flavor $\beta$ at point $z^{\prime}$,
is not diagonal, i.e. not zero for $\beta = \alpha$.
The free fields $\psi_i(z,t)$ can be quantized in the usual way by rewriting
the momentum distributions $b^s_i(p_z)$ and $d^{s*}_i(\mi p_z)$ in Eq.(\ref{23})
as creation and annihilation operators
${\sc B}^s_i(p_z)$ and ${\sc D}^{s\dagger}_i(\mi p_z)$.
The interacting fields are then given by
\begin{eqnarray}
\psi_{\sigma}(z,t)&=& \mbox{$\int_{_{\infm}}^{^{\infp}}\frac{dp_z}{2\pi} \exp{[i p_z z]}$}
\sum_{s=1,2}\{{\sc B}^s_{\sigma}(p_z; t)\,u^s_{\sigma}(p_z; t)\nonumber\\
&&~~~~~~~~~~            + {\sc D}^{s*}_{\sigma}(\mi p_z; t)\,v^s_{\sigma}(\mi p_z; t)\}
\label{A22}
\end{eqnarray}
where the new flavor creation and annihilation operators which satisfy canonical
anticommutation relations are defined by means of Bogoliubov
transformations \cite{Bla03B} as
\begin{equation}
{\sc B}^s_{\sigma}(p_z; t) = \mathcal{G}^{\mi \1}(\theta; t)\,{\sc B}^s_i(p_z)\,\mathcal{G}(\theta; t)
\nonumber
\end{equation}
and
\begin{equation}
{\sc D}^s_{\sigma}(\mi p_z; t) = \mathcal{G}^{\mi \1}(\theta; t)\,{\sc D}^s_i(\mi p_z)\,\mathcal{G}(\theta; t)
\nonumber
\end{equation}
By following the BV prescription
\cite{Bla95}, which takes into account the above definitions, it
was demonstrated \cite{Bla98} that the flavor conversion
formula can be written as
\begin{eqnarray}
P(\mbox{\boldmath$\nu_\alpha$}\rightarrow\mbox{\boldmath$\nu_\beta$};t)
            &=& \left|\left\{{\sc B}^s_{\beta}(\bar{p}; t),\,{\sc B}^s_{\alpha}(\bar{p}; t)\right\}\right|^{\2}\nonumber\\
            &&~~+ \left|\left\{{\sc D}^s_{\beta}(\mi\bar{p}; t), \,{\sc B}^s_{\alpha}(\bar{p}; t)    \right\}\right|^{\2}\nonumber\\
\label{A25}
\end{eqnarray}
which is calculated without considering the
localization conditions imposed by wave packets, i. e. by assuming that $p_z \approx \bar{p}$.
When the explicit form of the flavor annihilation and creation
operators are substituted in Eq.(\ref{A25}), it was also demonstrated \cite{Bla03B} that
the flavor oscillation formula becomes
\begin{eqnarray}
\lefteqn{P(\mbox{\boldmath$\nu_\alpha$}\rightarrow\mbox{\boldmath$\nu_\beta$};t) =}\nonumber\\
            && \mbox{$\frac{\sin^{\2}{[2\theta]}}{2}$}\left\{\left(1 -  F(\bar{p})\right) \cos{[\Delta E(\bar{p}) t]}\right. \nonumber\\
            &&\left.~~~~~~~~~~~~~~~~~~~~~~~+ F(\bar{p}) \cos{[2\bar{E}(\bar{p}) t]}\right\}\nonumber\\
            &\approx& \mbox{$\sin^{\2}{[2\theta]}
            \left\{\left[1 - \left(\frac{\Delta m}{2\bar{p}}\right)^{\2}\right]\sin^{\2}{\left[\frac{\Delta m^{\2}}{4 \bar{p}}t\right]}\right.$}\nonumber\\
            &&\mbox{$\left.~~~~~~~~+ \left(\frac{\Delta m}{2\bar{p}}\right)^{\2} \sin^{\2}{\left[\bar{p}t\left(1 + \frac{m_{\1}^{\2} + m_{\2}^{\2}}{4 \bar{p}^{\2}}\right)\right]}\right\}~~$}
            \label{A26}
\end{eqnarray}
where the last approximation takes place in the relativistic limit $\bar{p} \gg \sqrt{m_{\1} m_{\2}}$.
After some simple mathematical manipulations,
the Eq.(\ref{A26}) gives exactly
the oscillation probability $P_{\Dirac}(\mbox{\boldmath$\nu_\alpha$}\rightarrow\mbox{\boldmath$\nu_\beta$};L)$
calculated from Eq.(\ref{A14}) when it is assumed that the wave packet width $a$ tends to infinity and $t \sim L$.

This new oscillation formula tends to the standard one
(\ref{0000}) in the ultra-relativistic limit. If the mass eigenstates
were nearly degenerate, we could have focused on the case of a
nonrelativistic oscillating particle having {\em very} distinct mass
eigenstates. Under these conditions, the quantum theory of
measurement says that interference vanishes.
Therefore, as we have already appointed, the
effects are, under realistic conditions, far from
observable. Besides, in spite of working on a QFT framework, the lack
of observability conditions must be overcome by implementing {\em
external} wave packets, i. e. by calculating the
explicit form of Eq.(\ref{A17}) for fermions.
Such a procedure was applied by Beuthe for scalar particles \cite{Beu03} and,
in a very particular analysis,
with basis on the BV calculations and on
our {\em intermediate} wave packet results, it could be extended to
the fermionic case.

\section{Conclusions}

In this paper we have computed the modifications to the flavor
conversion probability caused by the introduction of the spinorial
form of neutrino wave functions.
To describe the time evolution of the mass eigenstates, we have
introduced wave packets constructed by superposing the Dirac equation solutions.
By following an analytical study with {\em gaussian} wave packets
we have computed the new effects that can be observed in the flavor
conversion probability formula.
Our study leads to the conclusion that the {\em fermionic} nature of
the particles and the
interference between positive and negative frequency components of
mass eigenstate wave packets modify the standard oscillation
probability which is obtained by implicitly assuming a {\em scalar}
nature of
the mass eigenstates.
Nevertheless, under particular assumptions, i.e. ultra-relativistic
particles and sharply peaked momentum
distributions, these modifications introduce correction factors
proportional to $\left(\frac{\Delta m}{2\bar{p}}\right)^{\2}$ which
are negligible in the ultra-relativistic limit.

We know, however, that the most rigorous treatment of oscillations
might be done in the quantum field theory framework.
Meanwhile, the prescription of oscillating neutrinos as Dirac spinors
was not yet completely and accurately described in a quantum field formalism.
The  BV model \cite{Bla95,Bla03} to neutrino/particle mixing and
oscillations is the most preeminent trying to this aim.
They have attempt to define a Fock space of weak eigenstates and
to derive a nonperturbative oscillation formula.
Flavor creation and annihilation operators, satisfying canonical
(anti)comutation relations, are defined by means of Bogoliubov transformations.
As a result, new oscillation formulas are obtained for fermions and
bosons, with the oscillation frequency depending not only on the
difference but also on the sum of the energies of the different mass eigenstates.

By using Dirac wave packets, we have reproduced an oscillation
probability formula with the same mathematical structure as those
obtained in the BV model \cite{Bla95,Bla03} in a QFT framework.
The study with Dirac wave packets enables us to quantify separately
each new effect present in the oscillation formula.
Imposing the initial constraint where we have a {\em pure} flavor-eigenstate
at time of creation $t = 0$ for any constant spinor $w$, we could calculate
the contribution of new effects to the oscillation probability.
Particularly, we have noticed that a term of very high oscillation frequency
depending on the sum of energies
introduces a very small modification in the characteristic of the oscillation
phenomena.
In addition, the spinorial form of the wave functions subtly modifies the
coefficients of the oscillating terms in the flavor conversion formula.

To conclude, we emphasize one more conceptual aspect arising from
the Dirac formalism. Dirac wave packets enable us to develop a
study of chiral oscillations \cite{DeL98}. In the standard model
of flavor-changing interactions, neutrinos with positive chirality
are decoupled from the neutrino absorbing charged weak currents.
In the ultra-relativistic limit, a state with {\em left-handed}
helicity is practically a state with negative chirality. If the
interactions at the source and detector are chiral, only the
component with negative chirality contributes to the propagation.
Therefore, the possibility of chiral oscillations can subtly
modify the oscillation formula. In this context, the study of
chiral and flavor oscillations could also deserve some further
specific studies.

\section*{Acknowledgement}

\begin{acknowledgement}

The authors thank the University of Lecce for the hospitality and
the CAPES (A.E.B) and FAEP (S.D.L) for financial support.

\end{acknowledgement}

\end{document}